\documentclass{cernrep}
\usepackage{lineno}
\begin{document}
\title{Search for Extensive Photon Cascades with the Cosmic-Ray Extremely Distributed Observatory}
\author{P. Homola\thanks{Institute of Nuclear Physics Polish Academy of Sciences, Krak\'ow, Poland} 
for the CREDO Collaboration\thanks{Full author list: http://credo.science/publications}}


\begin{abstract}

Although the photon structure is most efficiently studied with the accelerator instruments, there is also
a scientifically complementary potential in investigations on photons produced in the outer space.
This potential is already being explored with gamma ray telescopes, ultra-high energy cosmic ray observatories
and, since very recently, by the Cosmic-Ray Extremely Distributed Observatory (CREDO). 
Unlike the former instruments focused on detection of single photons, CREDO aims at the detection
of cascades (ensembles) of photons originating even at astrophysical distances. 
If at least a part of such a cascade reaches Earth, it might produce a unique pattern 
composed of a number of air showers observable by an appropriately dense array of standard detectors. 
If the energies of air showers constituting the pattern are relatively low and if the typical 
distances between the neighbors are large, the ensemble character of the whole phenomenon
might remain uncovered, unless the CREDO strategy is implemented.

\end{abstract}


\maketitle

\section{Introduction}
\label{sec:intro}
While the photon structure and it properties in the high energy regime are
investigated mainly using the data from man-made accelerators, one should remain aware of 
the scientific potential of astrophysical studies: gamma-ray astronomy and ultra-high energy cosmic ray (UHECR) 
research. Within the former field we deal with photons at energies unavailable in terrestrial instruments which adds
complementarity to the accelerator photon investigations, despite incomparably low flux of gamma rays reaching the Earth.
Considering the photon energies even larger than in gamma-ray astronomy, one enters the realm of UHECR which concerns
particles of energies exceeding 10$^{18}$~eV, with the few extreme events clearly above 10$^{20}$~eV. The existence
of such extremely energetic particles have remained a puzzle since decades. 
Interestingly, the two main classes of scenarios describing production of UHECR: 
``bottom-up'' models based on acceleration of nuclei and ``top-down'' class postulating stable existence 
and decay or annihilation of super-massive particles of energies reaching even 10$^{23}$~eV,
predict that photons should contribute to the UHECR flux \cite{Bhattacharjee:1998qc}. 
A clear distinction between the two classes is based
on the scale of this contribution: in ``bottom-up'' models one would expect very small fraction of photons 
in the UHECR flux while in ``top-down'' scenarios the photon contribution to the observed flux is
expected to exceed even 50\% at 10$^{20}$~eV. The research performed over the last decade by the largest
cosmic-ray instruments does not indicate the existence of UHE photons, thus stringent upper limits are placed
which under some basic assumptions might allow a severe constraining of the ``top-down'' class 
as a whole (see e.g. \cite{auger-diffuse-photon-2017}). The point that we undertake in this paper is based on 
a trivial note concerning the mentioned ``basic assumptions'': we do not know the physics at UHE, relaying on 
extrapolations over many orders of magnitude from the accelerator energy region. Being aware of the 
fundamental theoretical uncertainties involved in the interpretation of the UHECR data allows 
considering the variety of logical and observational consequences of taking significantly different 
theoretical assumptions. In this paper we highlight one of such consequences: since UHE photons are
expected to exist and the available evidence does not confirm their existence, we propose considering
scenarios in which UHE photons exist but have negligibly little chance to reach Earth due to the interactions
during their propagation on the way to us. Such scenarios prompt
a purely technical challenge: can one see the products of these interactions - ensembles of cosmic rays?
Actually this question needs an answer also within the state-of-the-art set of assumptions: if UHE photons exist, 
they should interact with the matter and fields during their propagation through the Universe which would lead to 
the initiation of extremely large cascades composed mainly of photons. And, continuing the logics within the paradigm,
we also ask whether under some circumstances the possible sizes of such cascades might compensate a very small flux,
as pointed out by the stringent upper limits to UHE photons. In other words we ask whether the scenarios involving UHECR
photons can be tested more efficiently with the focus put on possible detection of photon cascades rather than single
particles, complementarily to the current state-of-the-art research. 
The photon cascade approach has been initiated only recently by the CREDO 
Collaboration~\cite{credo-web,credo-general-icrc2017} and the scope of the addressed issues defines
a wide physics program with long-term perspectives rather than a short term project.
The basic research channel proposed by CREDO is the experimental verification of the astrophysical
models where particle cascades are initiated, with the emphasis on photon ensembles. Such a verification
would be possible only if there is a chance to observe at least partly the products of primary particle 
(e.g. UHE photons) interactions and this chance should be determined for the scenarios to be verified
before proceeding with the experimental effort. Complementarily, we also propose
another type of investigation which we call ``fishing for unexpected physics''. This 
approach is oriented on hunting for 
clearly non-statistical excesses above the diffuse and random cosmic-ray background, or arrival time correlations
of air showers and single muons (or other secondary cosmic rays) in distant detectors, 
independently of the expectations from theoretical models.
In this paper we highlight the CREDO science case and instrumentation and analysis
strategies related to these two research channels: testing scenarios and fishing for unexpected.

\section{N$_{\rm{ATM}}$>1: mysterious air shower observations and generalized cosmic-ray research}
\label{sec:ngt1}

It seems to be not very well known within the cosmic-ray community, especially among the younger colleagues,
that there exist published reports
on multi-cosmic-ray events looking like footprints of ensembles of primary cosmic rays 
correlated in time~\cite{smith-sps-b-83,fegan-sps-d-83}.
The reports discuss a) a burst of air showers at estimated mean energy of $3\times10^{15}$~eV lasting 
5~minutes~\cite{smith-sps-b-83}, 
and b) unusual simultaneous increase in the cosmic-ray shower rate at two recording stations separated by 
250~km~\cite{fegan-sps-d-83}.
Both observations were taken in two independent experiments in 1981 and 1975, respectively, and were the only events
of their kinds seen during the lifetimes of both detecting systems. 
Other few hints of such possibly correlated cosmic-ray phenomena were 
seen by some small cosmic-ray experiments dotted around the world, 
such as a Swiss experiment that deployed four detector systems in Basel, Bern, Geneva and Le Locle, 
with a total enclosed area of around 5000 km$^2$ \cite{cern-06-global-cosmic}.
As proposed in Ref.~\cite{cern-06-global-cosmic}, a globally coordinated cosmic-ray detection and analysis effort seems
to be in place to verify whether the peculiar air shower observations carry any physical essence or maybe are 
just artifacts. The proposal concerned building small and cheap detectors which were planned to be installed in high
schools at different locations around the globe, then operated and maintained by the high school pupils and staff. The 
science case addressed the cascading processes initiated by nuclei, mainly the 
photodisintegration of high-energy cosmic-ray nuclei passing through the vicinity of the Sun, first proposed by 
N. M. Gerasimova and G. Zatsepin back in the 1950s. The challenge had been undertaken in several 
research centers across the world where scientists in cooperation with their educational partners 
established small size experiments and approached a global coordination. Insufficient funding together with  
deficit of enthusiasm among the participants which grew with the continuing lack of exciting observations gave
no scientifically meaningful outcome and led to reducing or closing up the activity in most of the involved high schools.

Now the idea of a global cosmic-ray research is being revoked in an enriched incarnation of CREDO with the 
following novelties:
\begin{enumerate}
\item The enriched CREDO science case includes photon cascades which addresses 
foundations of physics at the highest energies, allowing constraints on e.g. 
Lorentz invariance violation (LIV) \cite{Galaverni:2007tq}, 
QED nonlinearities \cite{maccione08}, space-time structure \cite{maccione-liv-spacetime-foam-2010} 
or the ``top-down'' UHECR scenarios \cite{Bhattacharjee:1998qc}, similarly as in the UHE photon search.
Furthermore, the generalization of the global approach allowing consideration of photon cascades 
changes the detection strategy. Photon cascades might contain even millions of particles, comparing to
few or at best several in case of nuclear cascading like the Gerasimova-Zatsepin effect. This implies the
necessity of implementing novel algorithms and triggers, and at the same time gives
promising perspectives for unique, global observable signatures enabling 
event-by-event identification of cosmic-ray ensembles.
\item CREDO points to the necessity of involving as many detectors as possible, regardless the
technical diversity or complexity of the whole network,
focusing in the first stage only at the timing of single events and particles, looking for excesses in time
windows of different scales. This approach addresses a very wide variety of instruments: satellites, 
stratospheric balloons, cosmic-ray arrays, 
fluorescence telescopes, radio air shower detectors, gamma-ray telescopes (see Ref.~\cite{credo-gamma-rays-icrc2017} for
the first multi-primary gamma ray study in the CREDO context),
neutrino observatories, accelerators, educational arrays, university laboratories,
high school detectors, popular pocket-size detectors and finally smartphones equipped with a detection app
and educational toys with simplest detectors. 
Such a global network can serve as a universal scenario tester: 
a subnetwork of detectors meeting the optimum requirements
of a specific scenario can be used and more detectors can be built ``on demand'' if scientifically justified
by the expectations of the tested scenario. At the same time the network can be used as a whole to fish for 
unexpected physics, i.e. unusual rate excesses or arrival time orders, as highlighted in the Introduction.
\item As the acquisition of the data from ``everywhere'' recorded with ``everything'' would
generate an enormous stream of information, a sensible analysis and interpretation would automatically
require an enormous manpower, including an effort of non-professional but enthusiastic scientific partners.
Therefore CREDO takes the public engagement as the key scientific tool, putting the emphasis on the clarity of 
the key objectives, and easy, intuitive usage of the relevant tools. In parallel we will
offer the paths for both education and science careers
for all the participants.
\item CREDO puts a particular emphasis on the exploration of an alerting potential of the
global cosmic-ray network, addressing not only the astrophysical strategies but also
multidisciplinary research involving climate changes or seismic studies \cite{credo-general-icrc2017}. 
\end{enumerate}
With the novel approach to a global cosmic-ray effort proposed by CREDO it becomes evident that 
a) widening of  the scientific perspectives, b) including 
as much of the available data as possible, and c) involving as many of the potentially interested and enthusiastic 
colleagues as possible,
increases the chances for scientific discoveries of a fundamental importance. Therefore
CREDO postulates a fully open project, with free access to data and open source tools, where both financial 
and in-kind contributions are welcome. It all offers an unprecedented chance for multidisciplinary research and 
education based on cosmic ray data which are available everywhere and at negligibly small cost.

\begin{figure}[ht]
\begin{center}
\includegraphics[width=0.6\linewidth]{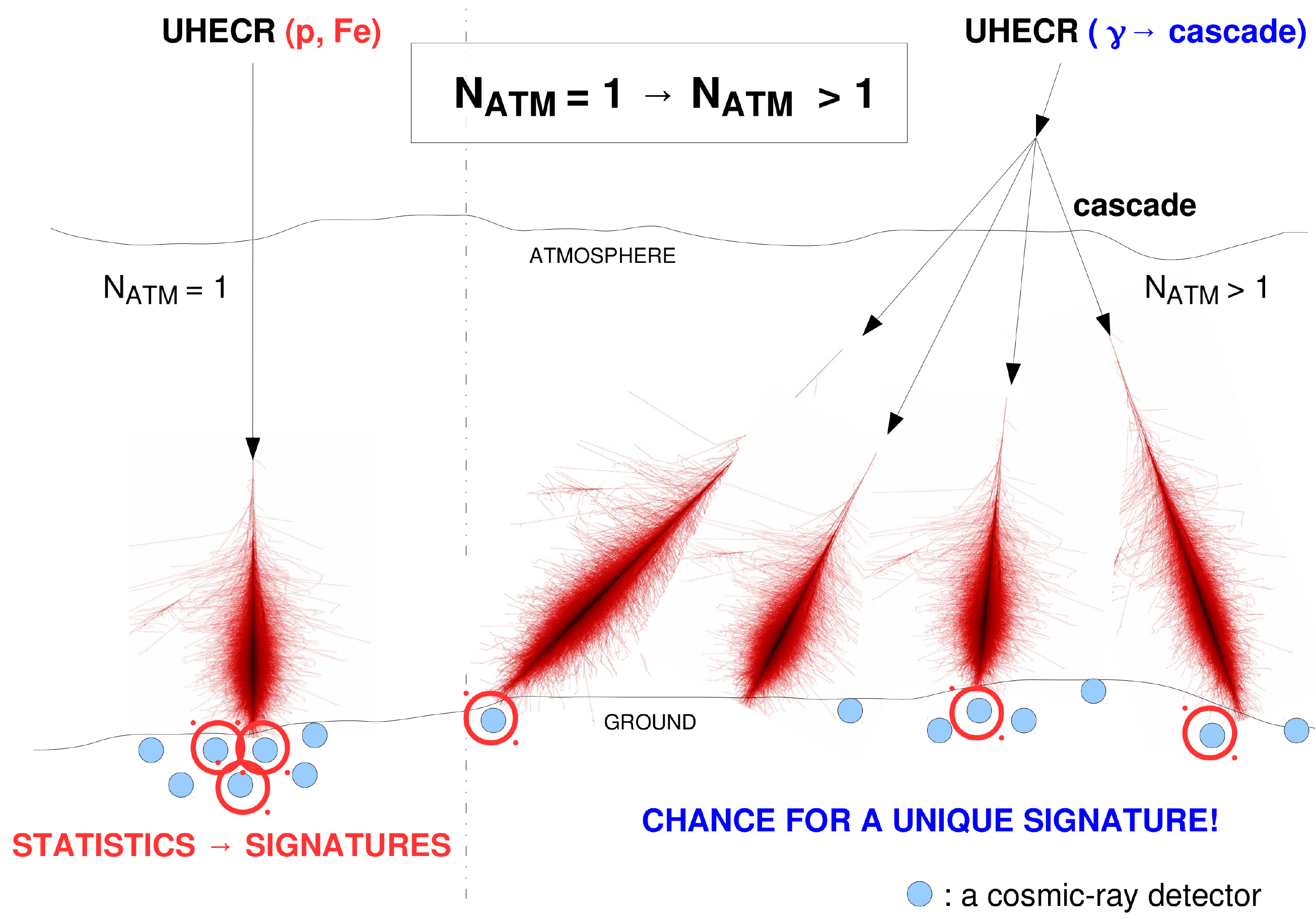}
\caption{Generalization of cosmic-ray research by admitting ensembles of particles as observation target.}
\label{fig:ngt1}
\end{center}
\end{figure}

Following the Introduction and the above considerations one defines the CREDO mission in the simplest way
by admitting more than one cosmic ray particle 
(including photons) entering the atmosphere simultaneously, where the term ``simultaneously'' denotes 
a temporal correlation and the specification ``more than one'' is equivalent to ``ensemble'' or
to the mathematical expression N$_{\rm{ATM}}$>1 (see Fig.~\ref{fig:ngt1}).

\section{Ensembles of photons: scenarios plus fishing}
\label{sec:ensembles}

The ``N$_{\rm{ATM}}$>1'' definition brings us to the already mentioned detection channels: 
A) testing theoretical scenarios and B) hunting for 
unexpected physics manifestations. Let us illustrate the channel A) with the two examples: exotic and standard.
The exotic example is based on noting that there are variants of LIV with 
critically different predictions concerning the UHE photon fluxes, depending significantly 
on the assumed alteration of the dispersion relation at the highest
energies. E.g. taking
the dispersion relation in the shape defined in Ref.~\cite{Klinkhamer:2008ky}:
\begin{equation}
\label{Eq-dispersion}
E_{\gamma}(\vec{k})=\sqrt{\frac{(1-\kappa)}{(1+\kappa)}}|\vec{k}|
\end{equation}
one understands that the sign of parameter $\kappa$ changes the UHE photon flux 
expectations dramatically. If $\kappa$ is positive, 
the pair production by
a primary UHE photon is suppressed, which should lead to increased UHE photon fluxes observed at Earth, in comparison to 
the implications concluded with using non-altered dispersion relation. In this scenario the non-observation result 
allows constraining $\kappa$ and therefore also LIV.
On the other hand, if $\kappa$ is negative, the lifetime of a UHE photon would be extremely short, 
even of the order of 1 second,
which on astrophysical scale is equivalent to an immediate decay 
\cite{chadha83-phot-decay,kostelecky2002-phot-decay,jacobson-2005-liv-rev}.
We note that
if the latter scenario is real, non-observation of UHE photons at Earth and the subsequent upper limits
would be a trivial, inconclusive result. However, even then one still has at hand one yet not checked research option -- 
approaching an observation
of products of a UHE photon decay: cosmogenic electromagnetic cascades. Although it is widely assumed that 
such cascades get completely dissipated before reaching Earth, thus contributing to the diffuse photon flux, there
are no precise calculations of the horizon (the distance within which a cascade can reach Earth at least in part, i.e. 
as an ensemble of a minimum two particles) with different theoretical assumptions. 
Such calculations within the Standard Model 
of particles are possible with the currently 
available tools \cite{crpropa2016} and the first steps in this direction have already been made, as described 
in Ref.~\cite{auger-targeted-photon-2017}. 
In addition, when one takes into consideration physics beyond the Standard Model, either of particles or cosmological 
-- more scenarios allowing observation of cascade-like signatures at the Earth appear verifiable
(see e.g. Ref. \cite{jacobson-2005-liv-rev} for a review on concepts relating to potential 
observation of quantum gravity manifestations). 
In this context it becomes apparent
that a complete study and search for UHE photons should include both an effort towards identification
of single UHE particles and a search for products of their decay: ensembles of photons correlated in time, most likely
dispersed significantly in space, maybe also in time, with energies spanning even a very wide spectrum.
The existence of a logically obvious and experimentally available, although yet not probed UHE photon search 
direction can be illustrated with considering two extreme cases: obvious detection of a photon ensemble
and its obvious extinction.
If photons in a cosmic-ray ensemble which reaches Earth travel very closely to each other, both in space and time,
they would induce a set of extensive air showers (EAS) which would effectively behave as one big EAS, being 
a superposition of the smaller ones, detectable with the state-of-the art techniques, e.g. with
a giant array of particle counters or with fluorescence telescopes. On the other hand, if the ensemble components
are distant one from another on average more that the size of the Earth, then obviously no conclusion about
the cascade-like nature of the phenomenon is possible: we see at best one particle which contributes 
to the diffuse and random cosmic-ray particle background. What is in between of these two ``extremes'', ensembles
of particles (photons) distant one from another on average by less than the size of the Earth, remains to be 
studied, and, possibly, observed.

\begin{figure}[ht]
\begin{center}
\includegraphics[width=1.0\linewidth]{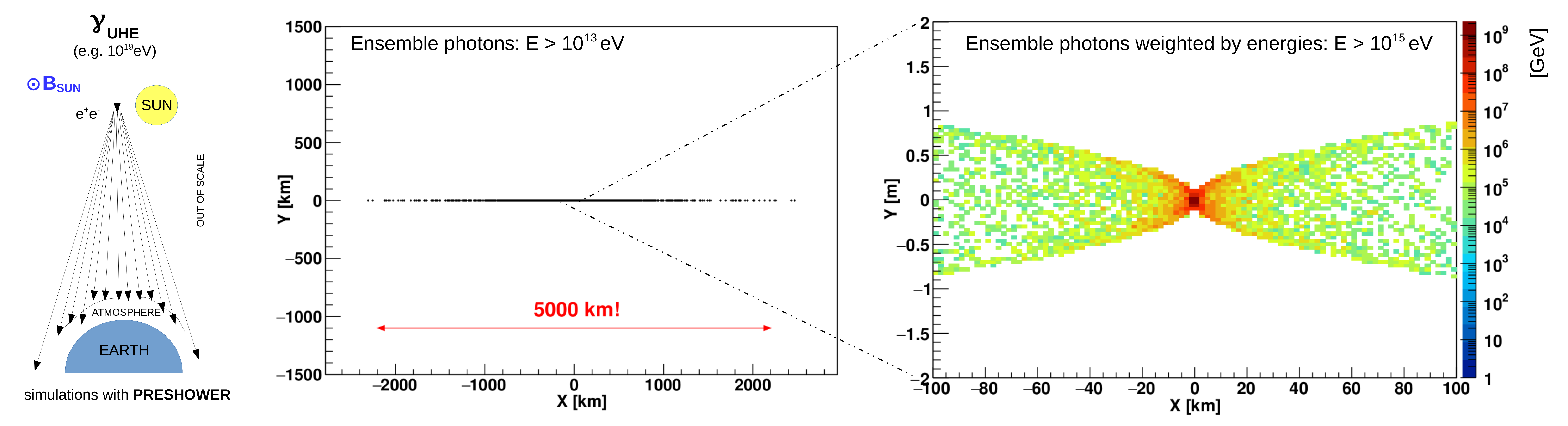}
\caption{The particle distribution on Earth of an ensemble of photons originated from an interaction of
UHE photon of energy 10$^{19}$~eV with the Sun magnetic field.}
\label{fig:sus-sps}
\end{center}
\end{figure}

An example of a non-exotic scenario within the channel~A is a cascade of photons initiated by a UHE photon 
primary passing through the vicinity of the Sun and interacting with its magnetic field (see Fig.~\ref{fig:sus-sps}).
This phenomenon, known in the literature as the preshower effect~\cite{erber66,presh-mcbreen81}, 
is expected within the standard 
quantum electrodynamics and can be simulated with the available open source tools~\cite{cpc1,cpc2} which
are also being used as a standard in the studies involving UHE photon-induced EAS~\cite{corsika}.
The expected particle distribution at the top of atmosphere is very much elongated (even 10000~km!) in the West-East 
direction and super-thin (meters) along the North-South line, promising a unique observable signature built of temporal
sequence of arrival times of the secondary cosmic rays on ground, and a very characteristic pattern of the triggering
detectors~\cite{credo-general-icrc2017}. In the preshower effect, once the primary UHE 
photon converts into an electron-positron 
pair, the electrons begin to radiate magnetic bremsstrahlung photons. The further the electrons travel
the lower their energies and the larger deflection with respect to the primary direction. This is reflected in 
the photon distribution on ground: the photons near the core corresponding to the primary direction
posses high energies as they were
emitted right after the electron-positron pair creation, when the electrons still had energies comparable to the primary
and they did not get deflected significantly in the magnetic field of the Sun. The further from the core, the lower photon
energies. In the example shown in Fig.~{\ref{fig:sus-sps} the primary photon energy is 10$^{19}$~eV and the 
spectrum of photons at the top of 
the Earth atmosphere extends from below GeV to above EeV (not the whole spectrum shown!). 
A feature such as shown in Fig.~\ref{fig:sus-sps} could be observed with a global cosmic-ray network, or with
a single large cosmic-ray observatory, or with a dedicated experiment tuned to the particle densities expected on ground.
Testing this scenario, which we call Sun-SPS (SPS for Super Pre-Shower), is one of the first scientific 
tasks of the CREDO Collaboration. It is worthwhile to mention that the largest observatories are tuned to 
record EAS with energies typical for the very vicinity of the Sun-SPS core, landing at distances not further than few tens km, 
while the whole Sun-SPS footprint might be even 3 orders of magnitude longer. It points to the advantage of
the global and diversified approach to the available cosmic ray data implemented in CREDO, 
at least as far as testing the Sun-SPS scenario is considered.

A special attention in CREDO is put to the channel B: fishing for unexpected physics. The idea
of the ``unexpected physics'' trigger based on arrival time correlations and order in 
distant detecting stations is sketched in Fig.~\ref{fig:mtrigger}.
\begin{figure}[ht]
\begin{center}
\includegraphics[width=0.6\linewidth]{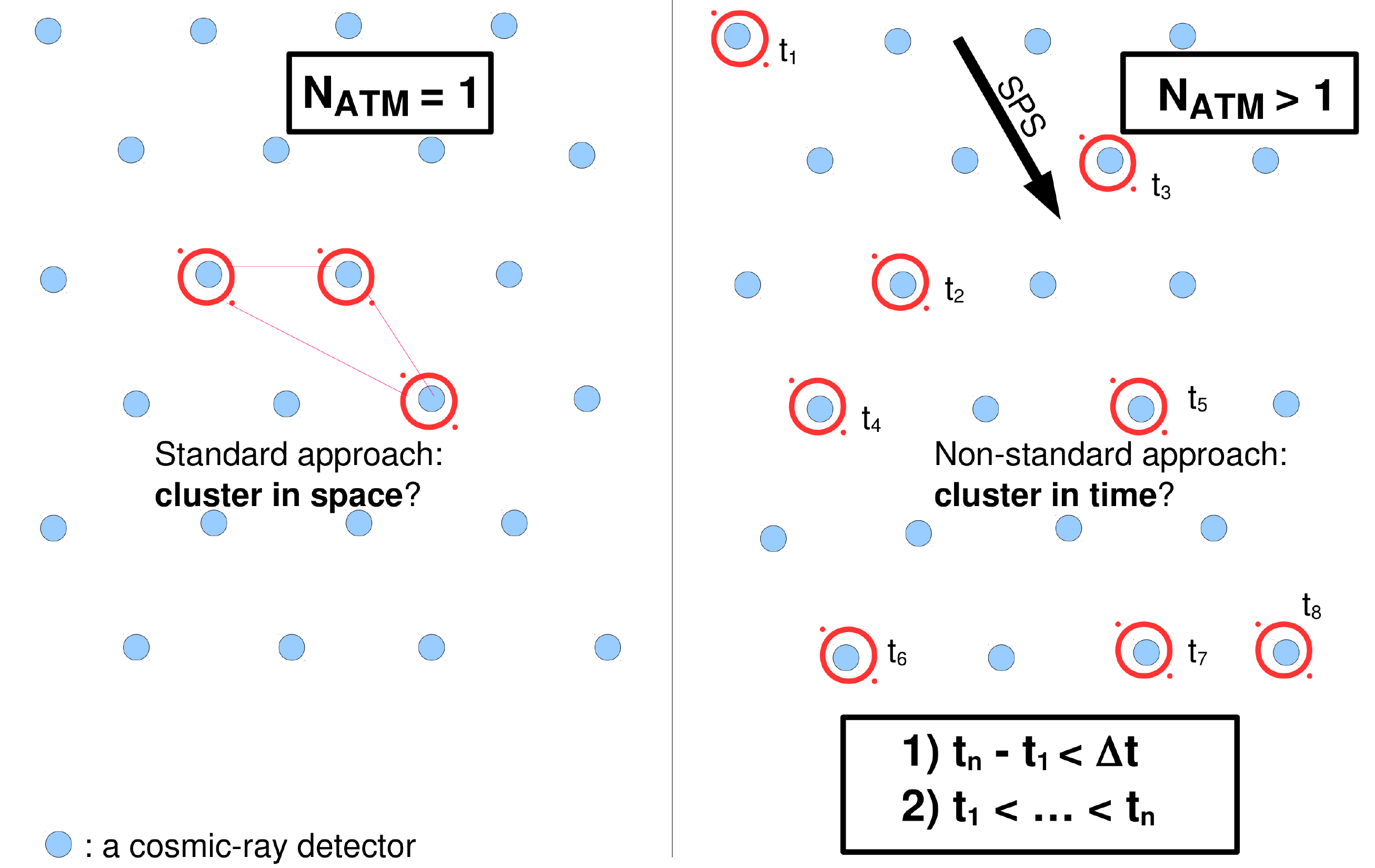}
\caption{A standard (cluster in space) vs ensemble (cluster in time) trigger in a ground array 
of a cosmic-ray observatory.}
\label{fig:mtrigger}
\end{center}
\end{figure}
Complementarily to the standard search of neighbor detectors triggered simultaneously by an EAS (clusters in space) 
one might also look for distant detectors triggered within some predefined time window by an ensemble of cosmic rays
(clusters in time). In addition one might expect some order in the arrival times of the particles or events
contributing to the cluster. 
The presence of such a feature would increase statistical significance of the observation.

\section{Observation: public engagement as a scientific tool}

As already explained, the scientific success of the CREDO mission is strictly determined by the
scale of the project possible to be achieved: 
total collecting area, geographical distribution of the detecting sites, and availability
of manpower. The optimum can be reached by combining the available professional resources and wide 
public engagement. Apart from social reasons for which the public should be kept informed and even involved in the 
professional scientific research, it is obvious that public engagement in an exciting 
scientific project must induce a growth 
of professional scientific resources bringing profits to the whole science community and to the society as a whole. 
The key condition for this scientific growth is to show
opportunities and paths of individual development and education within the project. In CREDO public engagement
is going to be driven by three simple tools that would help to reach both social and scientific objectives of the project.
Firstly, a massive participation will be achieved with an open source mobile application which turns a smartphone
into a particle detectors. Such applications already exist \cite{deco, crayfis} although they are not yet open, thus 
not enabling sufficient flexibility required for a society driven software engine. For this reason CREDO opens
its own app, to be freely distributed among science enthusiasts across the world with the encouragement to contributing
to the development \cite{credo-detector}. This of course does not exclude contributions from the users of the other 
applications to the common worldwide database. Another potentially available channel to involve even the youngest 
generations of science enthusiasts is related to the educational toys capable of detecting secondary cosmic rays 
and networked worldwide to help the CREDO mission. The detection of a particle and the link to the community 
dedicated to reach common and ambitious scientific goals should stimulate the passion, enthusiasm and a desire to get 
involved deeper in the project, i.e. to get educated.

Both using the smartphone particle detection
app or an educational toy will enable passive participation in the CREDO project 
by collecting the data. The next level of involvement
will be the activity within the CREDO community environment. The pilot component of this
environment is the CREDO citizen science platform Dark Universe Welcome (DUW)~\cite{duw} installed
on the Zooniverse engine \cite{zooniverse}. With the easily understandable analysis format of DUW (see Fig.~\ref{fig:map})
one will be able to analyze 
``private'' particles in the global context, 
search for ``strange'' detection patterns and help to train the ``scientific fishing'' algorithms. 
\begin{figure}[ht]
\begin{center}
\includegraphics[width=0.6\linewidth]{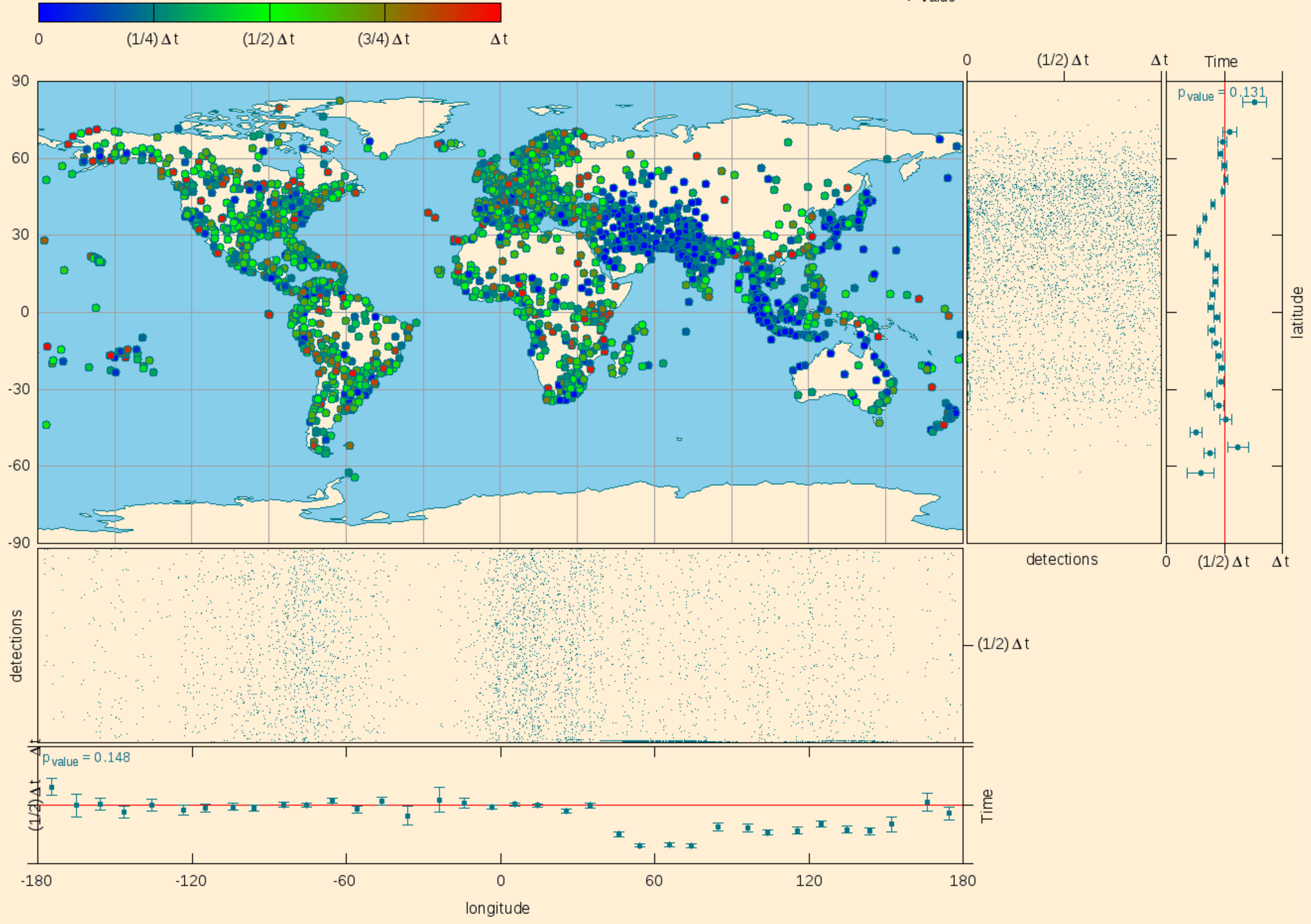}
\caption{A simple visualization of the ``fishing for unexpected'' strategy. The average arrival time 
within a certain temporal an spatial interval should be statistically consistent with the mean of the time 
interval if the received signal is composed of uncorrelated particles. A significant departure from the mean
might be a footprint of an ensemble of correlated cosmic rays.}
\label{fig:map}
\end{center}
\end{figure}
Other social facilities of the CREDO community environment, like e.g. individual and group rankings,
will increase the pleasure of doing science
and further stimulate the motivation to get involved deeper. The educational and scientific career paths supplementing
the popular devices and software will in turn strengthen the stream of creativity and ``fresh blood'' to power the
community of science professionals, that ultimately should be reflected in the increase of the ability of the 
society as a whole to develop by making scientific discoveries.  

The third and scientifically the most exciting tool of the public engagement will be the automated 
procedure to monitor the cosmic-ray data globally which will provide the easily classifiable monitoring images with 
the largest discovery potential. The prototype of such a monitoring machine, called CREDO Monitor, 
fed by some of the publicly available
cosmic-ray data, has been launched recently and is internally available for the CREDO members \cite{credo-monitor}.
The available data is migrated periodically from the active acquisition sites
to the data storage and computing center maintained at ACC Cyfronet AGH-UST~\cite{cyfronet}, then after
basic processing (scanning for time-clustering) classifiable global detector patterns (maps as in Fig.~\ref{fig:map}) 
are generated
and stored on a web server ready for the inspection with a human eye. 
The receivers of CREDO Monitor will be able to tune the view of the most interesting discovery proposals
selected by classifying machines and initiate a collective human-based classification according 
to the predefined crowdsourcing
requirements, and finally open a professional analysis with a variety of algorithms which would actually lead to
specifying of statistical significance of the proposed discovery patterns.

The above three pillars of the public engagement in CREDO should attract large number of participants and increase
the chances for a scientific success of the project. Importantly, all the contributions to 
data acquisition and analysis, no matter from scientists or from ``just'' science enthusiasts, would 
give the right to claim co-authorship of scientific publications and the share in the possibly accompanying awards.
Moreover, it is planned that the contributions will be easily registered and evaluated, leading to the estimate of 
the share in the project. Such an evaluation system, including e.g. the already mentioned
user rankings, would offer a potential to activate additional motivations of
the participants: positive competition.

\section{Summary}
\label{sec:summary}

We consider cosmic-ray cascades composed of photons correlated in time as a 
yet not checked channel of information about the Universe and the physics at the highest energies known.
If such cascades exist they might have a wide spatial distribution which might make them 
observable only with a worldwide network of detectors, and keep out of the reach of
even the largest cosmic-ray observatories with their state-of-the-art configuration. 
We introduce the Cosmic-Ray Extremely Distributed Observatory, the infrastructure and physics program tuned 
to cosmic-ray cascades, with potential impact on ultra-high energy astrophysics, the physics of 
fundamental particle interactions and cosmology, offering also a multidimensional interdisciplinary opportunities. 
We implement the CREDO strategy by applying a trivially novel approach to the cosmic-ray data taking -
a global and massive approach. 
Within the CREDO strategy based on the collective and global approach to the
available and future cosmic-ray data the chances for detecting and studying the astrophysical cascades
by definition exceed the capabilities of even the largest observatories and detectors working independently
of each other. Everybody, from theorists to non-experts, both institutions and private persons,
are invited and welcome to contribute.

\section*{Acknowledgements}
This research has been supported in part by PLGrid Infrastructure. We warmly thank the staff at ACC
Cyfronet AGH-UST for their always helpful supercomputing support.
The Dark Universe Welcome citizen science experiment was developed with the help of the 
ASTERICS Horizon2020 project. ASTERICS is a project supported by the European Commission 
Framework Programme Horizon 2020 Research and Innovation action under grant agreement n. 653477.
PH thanks Andrew Taylor, Marcus Niechciol, Daniel Kuempel, and David
d'Enterria for inspiring discussions.


\end{document}